# Mostly electric assisted airplanes (MEAP) for regional aviation: A South Asian perspective.

Vinamra Chaturvedi

Indian Institute of Technology, Kanpur, Uttar Pradesh 208016, India

email: cvinamra@iitk.ac.in

Dr. Vinamra Chaturvedi contributed to the concept, research, preparation, and presentation of this article.

**Abstract**

Aircraft manufacturing relies on pre-order bookings. The configuration of the *to be assembled aircraft* is fixed by the design assisted market surveys. The sensitivity of the supply chain to the market conditions, makes, the relationship between the product (aircraft) and the associated service (aviation), precarious. Traditional model to mitigate this risk to profitability rely on increasing the scales of operations. However, the emergence of new standards of air quality monitoring and insistence on the implementation, demands additional corrective measures. In the quest for a solution, this research commentary establishes a link, between the airport taxes and the nature of the transporting unit. It warns, that merely, increasing the number of mid haulage range aircrafts (MHA) in the fleet, may not be enough, to overcome this challenge. In a two-pronged approach, the communication proposes, the use of mostly electric assisted air planes, and small sized airports as the key to solving this complex problem. As a side-note the appropriateness of South Asian region, as a test-bed for MEAP based aircrafts is also investigated. The success of this the idea can be potentially extended, to any other aviation friendly region of the world.

# Introduction

Machine powered flight has an interesting history (**Petrescu, 2017**). A modern aircraft has undergone many changes in the design, over the years. Even though both modes of transportation are designed to carry weight, a surface-based vehicle (SBV), differs from an aircraft, in having a third degree of freedom in motion.

To completely track the motion of an aircraft, an altimeter is required. Altitude based mobility tracking is, an improvement, of the latitude and longitude based, ground position system (GPS). It was designed to track surface-based motion. The true position of an aircraft, in a volume of space, is detected, by a vector tracker. It is capable of locating, tracking and logging, the three variables (latitude, longitude and additionally the altitude) based position of an aircraft (**J. M. Eklund, 2005**).

A vector tracker has evolved from the initial days of paper-based record logging, to relaying the position-based statistics, by the use of internet protocol. Technological upgrades in these systems via the incorporation of fast electronic communications, has also helped in achieving autonomy in path-based motion control.

There are limits to the evolution of design of a surface-based mover to an aircraft (**Bejan, 2019**). Noticeably, the control of weight transfer, during the lift, float, altitude change and the landing actions of a flying unit, is different from an SBV.

For maneuvering at high speeds, without tipping, the center of gravity (COG) of the moving unit, needs to be kept as close to the surface as possible. This feature in the design of a moving unit is part of exercising greater control on the dynamical states of the vehicle. Physically, this involves monitoring and controlling the vibrations during motion. Shifts in the positions of the COG, changes the vibration characteristics and consequently, the stability of the moving unit. On a trip-to-trip basis, this can be gyroscopically logged and controlled (**Gross, 2009**). COG logs can be utilized to interpret the vibration signature, the performance and the standardization of the flight. This methodology helps in generating, a set of key performance indicators (KPI), by calibrating, against a set of known standards. It uses statistical tools, to create a framework, for the improvement of the transportation experience.

This approach has helped in the handling of the complexity of a machine powered flight, by providing a mechanism to intelligently control and improving the design of each iteration of the transportation unit.

Developmental activities in the transportation sector are a result of interrelated factors (**Kitamura, 1988**). Likewise, research in aviation centric transport, involves many variables. In this communication, the techno-commercial feasibility of augmenting, the fossil fuel-based power source (FFPS) of a mid-haulage aircraft (MHA), with electric power, is presented. Furthermore, the inter-linkages of the successful flights of such aircrafts, with the development of mid-sized airports, is studied, with the perspective of the growth of regional aviation.

A vibrant transportation network, ensures, robust growth of the economy. The phenomena could be regional, yet, its spread extends beyond geo-political boundaries. Indeed, vast time spans, of cyclical economic growth, experienced, by a substantial proportion of global population, were made possible by the developments of SBV (**Canning, 1993**). The second half of the last century, saw a rapid increase in the share of air-based mobility, alongside an overall increase in motorized transport (**Schafer, 2000**). The trend is expected to continue, provided the environmental sustenance is factored into the considerations of the developmental goals of the transportation sector (**Bakker, 2014**).

In comparison to the rest of the world, South Asia is endowed with a diverse and a different style of transportation network (**Bagler, 2008**). The fast-pace of the growth of the economy in South Asian region, in recent years, has been attributed to the rapid development in the transportation sector of the region (**Sahoo, 2012**). However, it is plagued by a rampant damage to the green forest cover, air pollution and general degradation of natural environment (**Hilboll, 2017**).

New infrastructure projects, related to transportation are the core-development area in the economic growth of regional geographies. Studies in the trends, of developmental economics, tend to indicate, that the historically high growth phases, can also be the times of severe or permanent damage to the environment (**Mebratu, 1998**). Thus, it is important to take care of the environmental considerations during the planning stage of any developmental project in the transportation sector (**McKinnon, 2015**).

Aspects of environmental order in air transport

The threat of worsening environmental conditions, could be due to

a) Poor implementation of the standards

and/or

b) Absence of environmental standards.

Either, any one or both the conditions are bad economics in the long run. Commercial gains from the improvements in air transport run the risk of getting, obscured, by the damaging and uncontrolled emission, from FFPS in automotive and aircraft engines (**Karen M, 2009**). Aviation generated sound, exposes a vast expanse of landmass to noise-pollution (**Hayes, 2014**). Timely interventions by way of product and service improvements, in the field of transportation, may help, in overcoming this developmental paradox (**Zanetti, 2016**).

Historic evidence, of the linkages of environmental degradation with increased fossil fuel use, from the already affected geographical regions, can serve as a model in planning, timely action-based interventions. The now vulnerable regions of the world, where the economic developmental needs are real, must simultaneously, be in a position to learn from the 'developmental mistakes' of the other regions.

Hence a goal oriented, learning model may serve well in these developmental effort. At the level of operations, the model may not be as linear as it appear. The differences in the choice of parameters, of the governing model must account for local factors (**Pucher, 2007**).

**The idea of mostly electric assisted propulsion (MEAP):**

One of the important factors in the development of sustainable regional air transport is the idea of using MEAP over FFPS for propulsion. Electric drives have been known to produce least vibrations, and were the natural choice, for vehicular propulsion in the initial years of motorized transport. However, the development of cost-effective filtering, of crude oil, helped in making the fossil fuel-based power source (FFPS), the preferred choice, for propulsion systems of the vehicles, in early part of the last century (**Hesla, 2009**).

What got overlooked in the societal rush for the adoption of motorized form of transport was the associated degradation of the environment. Noteworthy limitations, of the FFPS fossil fuel, are as follows:

Poor control of polluting effluents, from the source of power generation. It is now considered as the root causes, for the environmental disorder.

Even though, new reserves of crude oil are found at a rapid pace, the limitations of the filtering technology, reflects in the continuously increasing price of fossil-based fuel (**Deffeyes, 2008** ).

Yet the meaningful impact on the societal development, due to the use of FFPS based motorized transport is difficult to discount easily.

Competition to the FFPS, in certain class of vehicles, emerged from the idea of mostly electric assisted propulsion (MEAP). Till recently, the idea was considered far-fetched due to design limitations (**Schömann, 2015**). However, design proposals based on newly developed lightweight material for electric drives and batteries, offer hope of improvements, in payload handling capacity of MEAP based vehicular transport (**Peter, 2013**).

A breakup of the developmental approaches in the field of vehicular propulsion is shown in Fig.1.

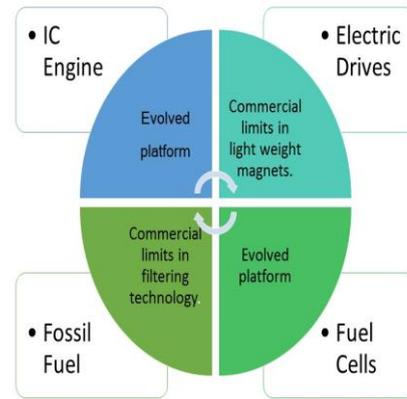

*Fig.1. Cyclic nature of developments in propulsion system derived from the design matrix indicating a towards Hybrid Electric Propulsion System for Air Transport.*

The figure indicates that the conversion of propulsion (CPS) is a techno-commercial problem. Further it conveys that the choice of the propulsion system, in a vehicle is limited by the selection-combination of the engine-drives and the fuel system.

Statistical tools are used in the generation of this design matrix (Fig.1) Such data-enabled linkages between,

a) The use of FFPS,

b) Design of transportation networks

And

c) Impact of a) and b) on the environmental.

These tools help, in the establishment of research idea in the field of electric vehicles (EV). The current state of art may not be sufficient, to establish a preference, from the available classes of haulage capacity, to be considered for the CPS project.

**Mid-haulage capacity as the best fit category for the conversion of propulsion system:**

Haulage capacity is defined as the weight drawing ability of the transportation unit (including the dead weight). There is a wide variety in FFPS, internal combustion (IC) engines used in SBVs and aircrafts. For an effective changeover to MEAP, in the existing and the new fleet of *in-vogue* transportation units, a haulage capacity linked class, must be identified. The, (re)fitment of electric drives, on this identified class of aircrafts; with fuel-cell based battery powered sources can be then be planned exercise.

Arguably, the long and the short haulage capacity aircrafts are equal contenders for the CPS. However, in this study only the mid- haulage capacity aircraft (MHA) is considered. There are several reasons for this choice. Some of them are listed below:

a) The used chasis and body-shell of a travelling unit can be of much use in a CPS project (**Mohaghegh, 2005**). The burden of unit-linked testing of the whole assembly of the aircraft can be reduced, by the proper implementation of the re-use technique. In doing this the performance of any new part, introduced in the travelling unit, can be tested under the simulated environment (**Smaoui, 2013**). Furthermore, the use of optimization techniques, such as regression, data of mid-haulage capacities can be extrapolated (interpolated) for large haualage (small haulage) capacities (**McCann, 2001**).

b) Weather related phenomena affect the dynamics of a flight. Flight tests in adverse conditions can use diverse models for the same (**Hauf, 2013**). Each flight disruption, during testing, gets added to the re-engineering effort. On a per unit basis, this cost is the least for the MHA category.

c) Stringent time bound test flights are required for the airworthiness, approval of a re-engineered aircraft. Material analysis of the parts used in the manufacturing of the aircraft, is the necessary condition, for the approval of test-flight (**Jones, 1995**). Frequent changes of the vulnerable parts, may also be required, at the testing stage. Tracing and replacement of the correct spare-part(s) for a stranded flight is a complex time critical exercise. Product-Lifetime-Management (PLM) technique is useful for parts replacement. An automated version of this technique for the aviation sector uses enterprise-wide resource planning (ERP) scenarios on networked computers (**Lee, 2008**). Running, a successful PLM-ERP system involves a three-party service contract between the customer, manufacturer, and the service provider. A transaction-based model, which uses "search" as the criteria of costing, is used in the design of these systems (**Asiedu, 1998**). In commercial operations, the cost of the search is shared between the stakeholders. For developmental efforts (such as the re-engineering of an MHA with CPS), numerous such searches from a variety of locations of the enterprise are expected. All this adds up in the developmental cost. It is necessary, to keep this indirect component of the developmental cost, of the re-engineered unit, to a minimum (**Rangan, 2005**). The spare parts belonging to the MHA category have the highest availability in warehouses, because of the design reuse philosophy, adapted by most aircraft manufacturers. Accordingly, the likelihood of finding spare-parts, for the MHA category, in the PLM-ERP system is the highest. Thus, in comparison to the other categories of haulage. The PLM-ERP search related component, of the cost of re-engineering with CPS, will be the least (**Benkard, 2000**).

Additional, considerations of safety, security, and compliance towards emission standards, also play a part in the choice of MHA for CPS (**Lu, 2006**).

## Use of Statistical tools in aircraft (re)engineering and aviation technology:

The two important variables used in aviation technology are related to the stability and control of the aircraft. (**Abzug, 1997**). For a conventional aircraft in flight, the aggregation of the possible set of moves are shown in terms of X-Y-Z (rectangular) co-ordinates defined as the Roll, the Yaw and the Pitch (shown in Fig. 2).

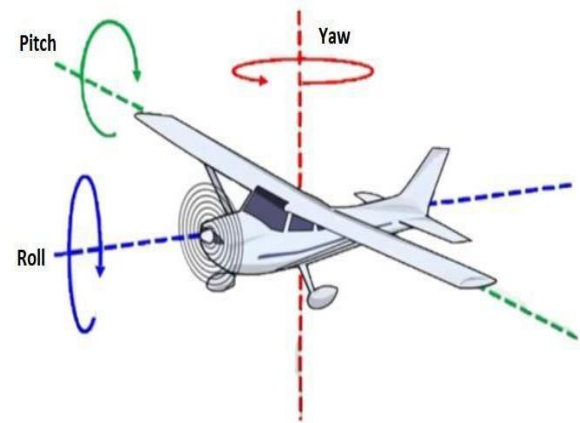

*Fig.2. Control of Roll, Yaw and Pitch based motion of an aircraft.*

In less conventional terms, a flights attitude and altitude need to be controlled to achieve stability of the transportation unit.

An active flights attitude and the altitude, can be measured and controlled from

a) within the moving unit,

and/or

b) outside of the moving unit

Following this a set of reference points can be generated to define a standard flight path.

Aggregated data-logs of the three-vector schema (i.e., the difference in altitude, displacement, and angular position, of the flying object from the flight path), can then be used for the real-time comparison of the (currently) active flights attitude. Statistical analysis of collected data such as mean deviation from the planned route etc. is done, to generate a knowledge base (KB).

This, plan can form the basis of vibration control in a MEAP assisted flight. Fixing the base level of noise signals to a new low, in each design-iteration of the MEAP aircraft unit, can be used to, calibrate and improve the transport experience (**Pickrel, 2002**).

Statistical analysis of big-datasets, result in the generation of a KB. Intelligence can then be drawn by the comparison of several such KB's, to assign key-performance-indicators (KPI), for the quantification of the flight experience (**Haghani, 2016**).

The generic testing, of the flying unit and the underlying philosophy for the preparation of the test-bed for MEAP based MHA is discussed, here:

A) **Vibration testing for the performance evaluation of the transporting unit.**

A network of sensors is formed by dispersing several such data logging units, on to the noise producing parts of the transporting unit (**Mieyeville, 2012**). Improvements in the product and the services in the transportation sector is based upon intelligent control of the vibrations from these parts. This approach, hinges on the reduction of unwanted noise (beyond threshold signal) during the movement of the vehicle. In manufacturing and testing of such vehicles, this methodology is used. Subsequently an intelligent controlling mechanism can be developed for the reduction of these vibrations (**McCall, 2004**).

Detailed time series analysis of the segregated dataset has also helped in the identification of the limits of a FFPS enabled transportation unit (**Samimy, 1996**). The per-unit trip-based cost, to the operator, is calculated based on adding the cost of maintenance, to the over-all cost elements of the functional (flying or motoring) hours. Since the maintenance cost can be co-related to the level of vibrations, a comparison of the cost of ownership of the FFPS based and MEAP enables aircraft can be done.

This, approach of comparing a set of external observables (from macroscopic data) of the flight with vibration logs (from within the flying unit) can be then be used to set/change the standards of testing (**Hu, 2003**). With the emergence of new standards for rich data centered communications, this operation can be done on a real time basis. Noticeably, the methodology of testing SBV's does not differ much from that of airplanes, in the statistics enabled regime. Accordingly, the design of a vehicle test bed (VTB) for air-based mobility (VTB*f*A) can be treated as an extension of a VTB for surface-based motion (VTB*f*S).

Based upon low vibration characteristics, electric drives have acquired popularity, for use in Electric and Hybrid-Electric Vehicles (EV & HEV) for some time now (**Zeraoulia, 2006**). A general understanding of the engineering principles indicates that a statistical correlation function can be established between the productive lifetime and the acceptable level of vibrations produced by the same unit during testing (**Eldred,1961**).

B) **Test Bed for the evaluation of flight with changed propulsion system.**

Air-transport is expected to follow suit of the developmental activities of EV and HEV of surface transport (**Zhang,2008**). Test flights of airplanes based on electric propulsion have been reported for some time now (**Romeo,2011**). Electric and hybrid electric airplanes are also a popular theme of recent research in electrical transportation (**Sarlioglu,2015**).

A test bed is required to test the flightworthiness of an aircraft. In view of the potential exigencies during a test flight of MEAP enabled aircraft, selection of the location for the VTB*f*A is important. Sites near soft surfaces such as forest canopies, water bodies, sand dunes and snow planes should be preferred. The location-selection surveys for the VTB*f*A can be based upon these requirements. Fig.3 depicts one such conceptual test bed for obstacle clearances. The entangled hard bed (inverted triangle in white in Fig.3), is the obstacle to be cleared by a flying object, while maintaining a prescribed flying altitude. The two sideward triangles (in blue) are the soft-beds, signifying, "safe-home", after the completion of in-flight maneuvers, over the hard bed-region.

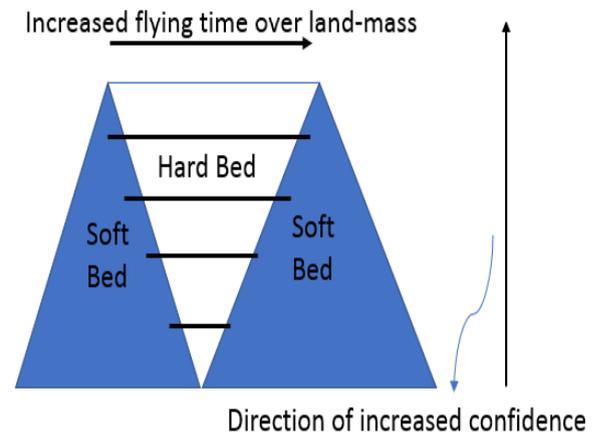

*Fig.3. Test bed concept with an inverse tri-angular hard bed (landmass) placed in between two tri-angular soft beds, water bodies, desert terrain, snow regions, forest canopy). Note the increase flying hours over land mass (x-direction), increases the confidence index (y-axis) in the technology used in flying.*

In the absence of standard procedures, it is important to have a broad understanding of the testing scenarios. Following features in the geographical area will help in the site selection of VTB*f*A :

1) It should cover for a large variation in the weather zones. (Ideally the short, the medium and the long-range of test flights)

2) Test area, should have "room" for sudden and/or smooth changes in the altitude of the flying object.
3) Sufficient proximity between soft-bed (water bodies, snow covers, desert areas, marshy land and/or thick forest canopies) and artificial or natural hard beds like asphalt cement tracks or meadows.

A sufficiently large variety of natural landscape formations in a medium sized peninsular land mass, with scattered and avoidable urban spaces, and high peaks of the mountains, represents, an ideal testbed for a MHA's test flight. Choice of a flying region with all the above attributes, over a land mass, which is free from human interventions, is important.

Peninsular and Central region of the Indian subcontinent, shown in Fig.4 which is populated with many sizeable water bodies, fits the aforementioned description. This region is also home to sand filled coastal area and offers a unique opportunity to test flights between seacoasts.

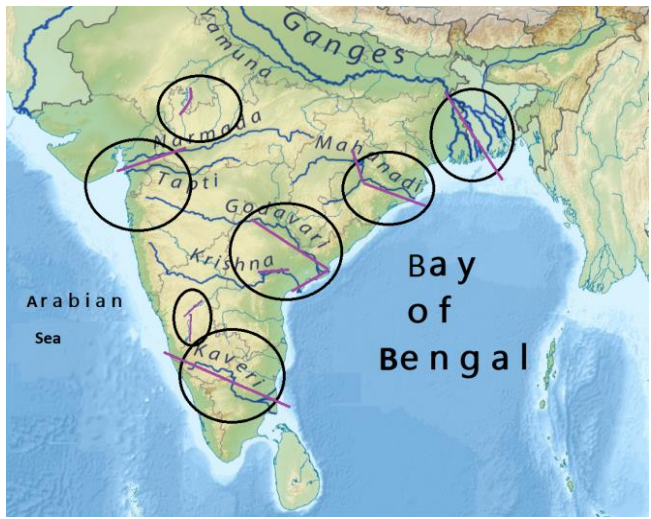

*Fig.4. South Asian peninsular region for the test bed development, around inland water bodies and seashores. Black circles represent testing areas of interest and purple lines show potential flight paths, while staying close to the water bodies (soft beds for emergency procedures during test flights).*

In an ideal test plan, the confidence-index (CI) of the success of the test-flight should increase simultaneously with an increase in the potential risks during the testing procedure. This requires continuous calibration of test conditions. In case of test-flights of re-engineered aircrafts, following provisions are required

i) Increased flight - time over land mass with sufficient provision for soft-bed landings in case of emergency.

ii) Identifiable soft beds (like seacoast, forest canopy, water catchments) in case of failure during landing, take off and other in-flight (hard maneuvers).

iii) Means to calibrate the CI

As is clear from the map of peninsular India (shown in Fig.2), almost all the provisions of the ideal test bed can be met by the circled and line-marked regions. Thus, this region may prove to be an ideal test bed for the mid-haul range flights of re-engineered hybrid electric aircrafts.

## Hybridization of the engine and the changes in the fuselage section for mid haul capacities:

Weight, of a transportation unit, is distributed between,

(1) Engine & related components.
(2) Passenger/cargo-section

And,

(3) Fuel tank.

The concentrated load of the engine, along with the continuously decreasing weight of the fuel tank, (used in the propulsion system), contribute a large portion to the deadweight of a vehicle.

Once the haulage capacity and transportation environment, is fixed, the cargo section of a vehicle can be designed independent of the engine ratings **(Husain,2011)**. Thus, in theoretical design, a railroad car can provide the same haulage capacity as that of a truck. However, in practice, the right ratio, of the volume and the surface space of the fuselage, engine, and cargo sections, of a vehicle, along with the requirement of maintaining a proper center of gravity, is required for the right design of a transportation unit. Keeping this view, any changes in the propulsion system of a (already designed) vehicle, will require an exhaustive re-testing plan. This way, the requirement for VTB$f$S and the VTB$f$A are inter-related.

The change(s) in the weight during high-thrust maneuvers causes an unprecedented change in the fuel demand. This makes it tougher to control an airborne vehicle, in comparison to a surface mover **(Asher, 2017)**. The effect of weather-related phenomena on the control and dynamics of the flight (discussed in the subsequent section) imposes an additional constraint on the design of an HEV aircraft.

Independent of the propulsion system, the requirement of controlling (or steering) the aircraft (or the SBV) remains the same for the successful execution of the trip. As shown in Fig. 1, previously, complete or partial control of the arrested motion within the three degrees of freedom is a complex exercise in control engineering **(Smetana, 2012)**. Off the total budgeted energy, for a planned trip, terrain specific issues can cause deviations from the budgeted fuel for the trip. Such deviations arise due to weather related changes on the trip dates or poor operator skills.

This data can then be used for the purpose of statistical analysis in the future. Data logs from test flights can be arranged according to the following table:

Table 1

| 2 ↓   1 → | On Board | Off - Board |
|---|---|---|
| In Flight | YES | YES |
| Off Flight | YES | YES |

*Table 1. An aggregated tabulation of the testing criteria, of a re-engineered aero-plane. Each of the four entries in the table can be further expanded into a child-table. The aggregated dataset help is easy identification of a complex problem.*

Results of the in-flight- testing of a re-engineered aircraft, may impose certain restrictions on specific maneuver. These limitations should reflect, in the modifications of the moving unit's controlling mechanism. Consequently, a re-registration exercise can be prescribed for the limited airworthiness of the unit. All this can be done within the policy framework of the air transportation administration authority.

## Design and testing of mid haul capacity vehicle on customized test beds

Some parallels can be drawn between the testing of an SBV and an aircraft. For example, in terms of testing the difficulty levels of hard tasks, an uphill maneuver, for an SBV, is equivalent to, an altitude climb, for an airplane. Similarly, a sharp turn during an uphill climb, for an SBV, is akin to an altitude climb in crosswind conditions for an aircraft. For such tough maneuvers, during the testing flight, close to maximum thrust is required. The surface and air vehicles are designed accordingly. The ratio (**R**) of the dead weight, to the dynamically changing weight (during the execution of the maneuver), is an important parameter. This factor is of importance, in the design of control mechanism of the aircraft (**Bradley, 2007**).

A successful design project for mobility, allows provisions for additional fuel requirements, for exceptional circumstances (**Fitzpatrick, 2006**). Much effort is required, to keep the value of '**R**' within the permitted range. In most cases this is done by reducing the deadweight (**B. P. Mann, 2012**). It can also be achieved by increasing the emergency provisions of fuel in an aircraft (**Raymer, 2012**).

In complex problem, such as this, a part-design-part-system-oriented approach can be taken. Sometimes this is considered as a better option, to arrive at an approximate solution (**Pinon, 2012**). Additionally, computer-based techniques for the design improvements can be deployed. In one such instance, much of the developmental effort of the aircraft design is reduced, by visualizing the aerodynamic behavior of the unit, inside a virtual "wind-tunnel". Experimental results can be compared with computer-generated models in such projects (**Bryson, 1992).**

Alongside of the "body and frame" of the aircraft, improvements in the propulsion system have been reported using this methodology (**Resende, 2004**).

Based upon, the position of the propellers, on the body, the design, can range between

a) the concentrated-load-centric (captive-engine - style), of a nose engine aircraft, to,

b) the distributed-load-style, of "over the wings and tails engine", aero-plane.

The two extreme points can serve planning, the (re) engineering effort of the MEAP based aircraft.

The basics of aircraft design, hinges on the haulage capacity. It is required, at the design-build stage of the manufacturing cycle (**Marks, 1973**). In practice, once the power requirement is fixed, (to some arbitrary range), a hybrid-electric propulsion design, can be approximated by a selection-combination of electric motors and/or FFPS. (**Filippone, 2000**).

As discussed previously, the dead weight of the combustion engine and/or the electric system influences the value of '**R**'. Several iterations of running the optimization problem (derived from the design matrix approach), is used to test and assign, a range of values to the haulage capacity of the aircraft (**Weisshaar, 1994**). Furthermore, this approach can also be utilized in the design and/or complete replacement, of the FFPS, based propulsion unit with a fuel cell powered MEAP system, after proper testing (**Emma Frosina, 2017**).

## Need for indigenous medium haul capacity for regional aviation in South Asia:

The sustainability of growth, of the aviation sector across the globe, comes from proper planning (**Ryerson, 2014**). In compliance with the globally accepted practice of subsidizing the taxes for a new airport development project and the airport fee, the same is implemented by regional planners. The limited impact of this on the overall development of regional aviation has been attributed to non-market driven factors (**Zhang A. H., 2008**).

A diverse demographic, seasonal nature of demand, fluctuating fuel price and large airport-usage tax, are some of the factors which affect the aviation-based transportation planning. In South Asia, air-transport is popular in areas, where roadways and rail-network is sparse (for e.g. coastal and hilly regions). The preferred ecosystem of transportation (buses, roads and bus stations, rail, railways, and rail stations) is unreliable. It faces frequent breakdowns, during peak demand season. This could be due to overcrowding and other factors. On the other hand, the popularity of aviation-based transport, remains limited, to long distance routes, even during, high peak season. The regional aviation factor in South Asian transportation network is minimal.

Following set of alternatives can be considered for a seamless integration of regional aviation in the ecosystem of South Asian transportation:

*A) South Asia specific need for mid-haul aircrafts*

Of the possible solutions to come out of this economic cycle lie in the technical as well as the commercial aspect of the aviation trade.

**Early inception of Airport projects:**

A new or an old airport project may have several stakeholders. This could include government and private agencies. Commissioning of airports in remote regions may require some form of incentive. In most cases it is in the e form of reduced tax rates on land use, at the time of the project initiation. Initial monetary flow from the pool of earnings, formed out of the operational activities of the other airports in the region can also be directed into the new projects. This mode of planning for a new airport requires fiscal discipline. Good understanding of the regional traffic dynamics will also help.

The stages of evolution of a new (or re-development of an old) airport project is shown in Fig.5.

Noteworthy points, in the figure are as follows:

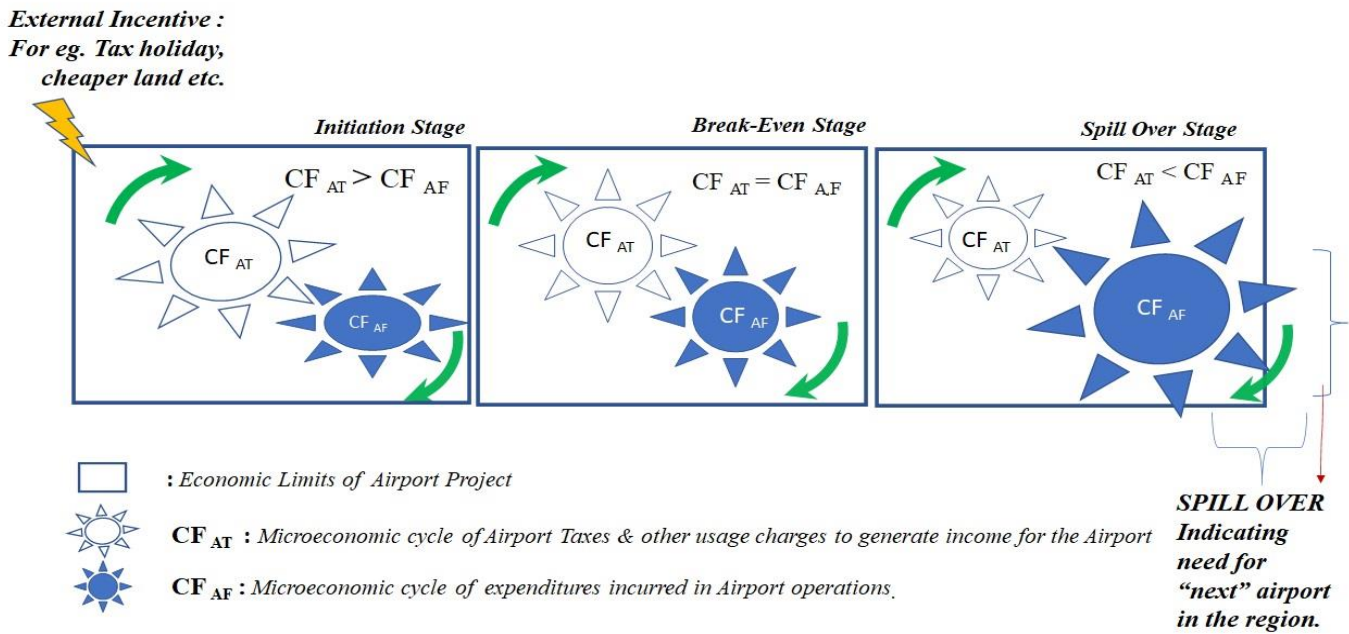

*Fig.5. Various states of existence of a regional airport, in accordance, to the stage of evolution of the services offered. Each state-stage can have an independent existence in the functional lifetime of the airport. Also, each state could be a dependent stage in the long-term evolution of the airport.*

**A) Technical Solution**

*i) Mid-size airports*

Populated areas in South Asia fall into the ancient to old category. Most airfields are relics of the colonial past, with rural road style of air strips. In the recent past, any announcement of modernization of these "airports" was followed by a swift and steady increase of the nearby real estate prices. By the time an airport project nears completion, the airport taxes, become the main source of sustained income **(Pitfield, 1981)**.

- A new or old air-port project requires a steady flow of cash, generated from the earnings and expenditure of the daily operations of the airport. An aviation-based transport project in the transportation ecosystem is propelled by several micro-economic cycles (µEC). Chief amongst these is the Income cycle from taxes and usage charges ($CF_{AT}$) and the Expenditure cycle of the operations ($CF_{AF}$).

- For the airport project to reach a state of optimized operations, all the µEC should balance each other.

- To start airport operations financial initiation is required (shown by yellow "bolting" mark, in the figure). Service Level of the operations is independent of the stage of evolution of the airport. For eg. An airport project can be in a state of very low usage, but from external incentives (like tax-breaks, fund-infusion etc.,), it can quickly acquire a stable state. From this current stage, the airport project, can "ex-gress" into a Spill-over stage. Secondly, it can, also "in-gress" back to the initiation stage. Thirdly, it can also continue to remain operational in the stable state.

- In the spillover stage an airport project starts losing money. This could be due to higher operational cost (shown in the third stage in the figure where $CF_{AT} < CF_{AF}$), lack of fund discipline, influx of un-chartered flights etc.

- For an optimal stage, in the operations of an airport project, the earnings and the expenditure cycles balance each other ($CF_{AT} = CF_{AF}$)

An increase or decrease in the traffic, results in a commensurate, increase or decrease of the μEC, respectively. This reflects in the scope definition of the airport project. The "blue-box" in Fig. 5, defines the size of the airport economy. Naturally, this adjusts, according to the "demand and supply" of the traffic needs.

However, in where the traffic pattern is random, further expansion of the airport project runs the risk of becoming un-economical. This is shown as the "spill-over" in Fig.5.

**The Regional Airport in the South Asian context:**

In the South Asian context, the identification of the stage (or state) of the airport (Fig.5) may become very difficult. From the perspective of managing the project, the starting phase may lie on a sparsely-populated, semi-urban tract of land, which is old. The choice of remote location is driven by security and safety issues of low flights, over densely populated areas. As the construction activity, gather's "momentum", social demographic of the site may change rapidly.

Most aviation regions in North-East Asia have airports, which started in remote locations but are now severely congested in terms of air-traffic and the peripheral real-estate (**Ha, 2013**). In recent times South Asia aviation region seems to have followed suit. In economic terms, the elasticity of demand is such that the "sources of income", run the risk of becoming "sinks of expenditure". The "inflection point" in supply and demand of aviation-centric traffic, is difficult to identify on a time scale. In the 'hyper-dynamic-ecosystem' of regional air transport, this may result in bad project planning.

Once incepted, the success of an airport project gets calibrated against usage (**Abdel Aziz, 2007**). Clearly, this approach has limitations. In certain cases, the time frame of the construction phase may out-stretch into the time frames of the implementation and launch phase (traffic-handholding phase). Consequent cost escalations, result, in the shift of focus of the project-deliverable, from a need to manage the air-traffic to that of increasing the traffic.

In most cases, this phenomenon is driven, by a need for loss-compensation via revenue generation. Once, a revenue-driven, supply chain builds, it is difficult, to shift the profit centric focus of the operations to other non-commercial developmental goals. Noticeably, the long-term commitments to environmental sustainability may suffer.

Airport expansion, in terms of size and traffic, seems to be an obvious choice for popular destinations. Yet, the long-term repercussions, of this choice, may not be the best for the development of an ecosystem of smaller hub-airports in the destination region. This philosophy of curtailing the immediate demands of operational readiness, by policy-based interventions, towards long term goals, may be very useful in the *true* development of regional aviation (**Postorino, 2010**).

Thus, the right approach may lie in "de-capitalizing" large sized airport project, after the "inflection points" are identified. Furthermore operationalizing, new and/or smaller airports of a region, may require, support from airline fleet operators. Hence, for a regional airport project to reach an early 'stable stage' (see Fig.5), increased use of mid-haulage flights could be a better option.

### ii. Mid-size aircrafts

Any 'last-minute" changes in the flight plan can result in a Flow Management Problem (FMP) (**Odoni, 1987**). Yet, this is an option offered by most air traffic controller and regulators (ATCR). In, where, aviation is the preferred mode of transportation, such changes in the flight plans, can be a regular feature.

To deal with FMP in aviation, automation (electronics, communications, and special software in networked computers) is required (**Gilbo, 1997**). Physically, it involves, de-congesting airspaces, around populated area. It also deals with a wide variation of weather data. Absence of reliable telecommunication infrastructure, in the early years of civil aviation, prevented this practice from becoming popular, amongst the network of South Asian ATCR's. However, the ability to handle the demand for 'last minute' changes in the flight plan (including the change of aircraft configuration), remains a good indicator of the airport's operational readiness (**Burghouwt, 2009**)

Increase in the number of MHA's, in the aviation fleet, is directly related, to the flexibility of operations. Growth in activity in regional aviation, because of increased interlinkages between mid-sized and small airports has also been reported (**Swan, 2002**).

Evidence of air quality deterioration due to increased use of regional airspace has been reported (**Masiol, 2014**). Any changes in the air transport policy, directed towards increasing mid-haulage aviation for regional aviation in South Asia, must also have provisions to mitigate these risks.

Thus, the evolution of MEAP technology for mid haulage flights, has the potential to be *truly* useful for the *true* development of aviation on a regional basis.

### C) Commercial Solution

True growth, in a regional sector of aviation comes from the increase in the number of destinations offered by operators. This in turn depends upon the development of new routes (**Heinz, 2013**). In open markets, the air passenger travel and cargo movement are tied to the sector wise growth of regional aviation (**Fu, 2010**).

The absence of mid-haul capacity aircrafts, in price-sensitive regions, such as South Asia, can expose the growth trajectory to negative economic trends. The lack of popularity of MHAs amongst airlines may also be linked to the "operational inertia".

***On the limitation in the current policy framework for the generation of developmental funds for new projects in the aviation sector:***

In the prevailing model of business, most South Asian airline operators, lease large sized aircrafts. A large portion of the cost of running operations of the airport is derived from the rental use of the airport space. In normal course of operations, the airline operators, tend to pay the airport charges to the airport authorities (AA). In-turn the same is recovered from the user as an indirect service. In most countries, this fund is managed by the airport authorities (AA). Most of it, gets utilized in the name of maintenance and expansion activities of the airports.

On occasions of increased monetary collection, in this fund, the need for seed capital for new projects in regional aviation (such as the redesign of electric grid for bulk battery charging, aircraft–engineering, etc.) can be planned. Since most such projects are funded under the public-partnership mode (PP), the balance of payments can be generated from the public sources.

Such flexibility is missing is the current level of operations of most regions in the world of aviation, due to reasons which are beyond the scope of this communication. Thus, in most cases, the commissioning of the new airport project, gets differed, till a regional (local) funding partner, is tapped. Excessive techno-commercial control by aviation authorities in certain aviation regions and lack of co-ordination between provincial or local government

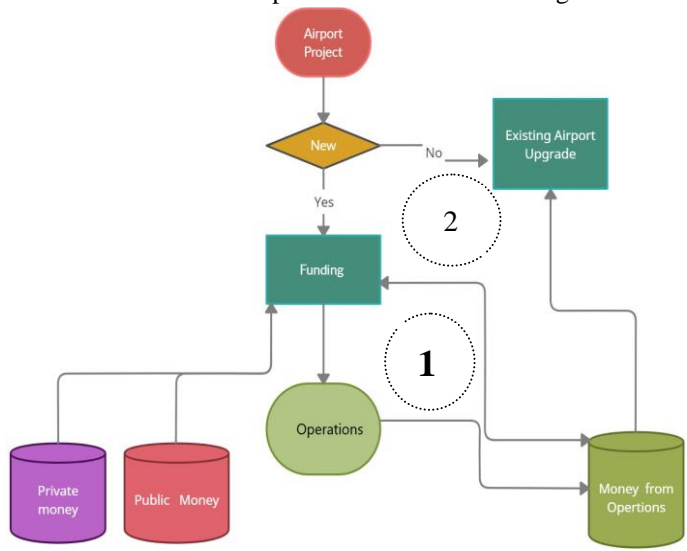

*Fig.6. A potential airport project, funding pattern, in flow chart format. Dash-circled (numbering 1&2), represent, quasi-static, micro-economic cycles (µEC), within the large macro-economic cycle.*

agencies are some of the reasons, why such developmental projects in regional aviation get delayed.

Irrespective of the delays, certain fundamentals in the (re) development of the airport projects remain the same. These are shown in flow chart in Fig.6. There are noticeable, self-fulfilling closed loops economic cycles (marked as 1 & 2 in Fig.6).

At an operational level, this signifies, what has already been noted in fig 5. Due to the imbalances of the micro economic cycles of the airport project, there is a tendency in the entire airport project to gravitate towards, the demands of serving the large aircrafts. Such events in the µEC of the airport project, obscures the *real*, (and possibly new) pattern of traffic demand (for e.g., the emerging requirement of mid-haulage aviation in the geographical vicinity of an existing large sized airport). This can result, in the complete localized exhaustion of excess revenue from the pool of collections of the existing airport. Net result of the phenomena is the emergence of **single** large airport in the region, **instead** of numerous small feeder airports.

Thus, it is important to develop a interventional policy framework to allow for the diversion of a fraction of the access fund, towards the development of a regional aviation. This could be in the form of a regional aviation development fund part of which gets used in the re-engineering of the mid-haulage aircraft. Another part of the same can be utilized in the developmental efforts of several mid or small sized airports in the region.

**Techno-commercial barriers in the manufacturing of mid-haul civil aircrafts in South Asia:**

The majority of mid-haulage aircraft manufacturing moved out of the United States to regional players in North and South America during the NAFTA regime. Some attribute this change as a blessing in disguise for the manufacturers in North and South American regions (**Morales, 2008**). In Asia, the phenomena of decentralization of aviation from government sponsored carriers happened simultaneous, to the events on the other side of the globe.

The laws governing the operations of airlines are also linked to the environmental protection laws. Old airplanes are permanently, de-registered due to strict compliance of the environmental standards. In Asia, where the interpretation and administration of aviation standards could be different, such deregistration could be temporary (**Thian, 2015**).

To compensate for the losses due to such "groundings" the airline operators, switch to bulk lease transfers of aircraft fleets from other regions of world. This practice curtails the demand for locally manufactured aircrafts to an artificial minimum.

**Aircraft manufacturing in South Asia**

The base year of aircraft manufacturing in South Asia, can be fixed to 1951. This is when trainer aircraft (HT Marut), was assembled in India. However, the manufacturing of MHA segment still faces a shortage of demand (due to reasons mentioned earlier in this text). Additionally, the suppliers-base for parts required for the development of regional MHA is missing, due to uncertainty in demand. Meanwhile, new manufacturing hubs in Brazil (Embraer) Canada (Bombardier) and Japan (Honda) have emerged. These hubs already cater to the medium haul range demand of the global aircraft industry.

**Aircraft re-engineering: Reasons and technique**

Active life of an aircraft is calculated on the basis of the number of pressurization and depressurization cycles, during operations. Indexed with haulage capacity and the engine lifetime, this translates into 25 to 30 years of active service life (**Brewer, 2013**). Based upon this, the aviation authorities, and regulators, run, air -worthiness certification programs.

Most airline operators participate, in manufacturer sponsored surveys, of the current state of the 'in-service' fleets (**Papakostas, 2010**). The prevailing practice in the aviation industry has been, to use this data to predict the future demand of the aircraft. This form of pre-order bookings helps in ensuring that the supply chain required in aircraft manufacturing remains active. Efforts in the development of MEAP based MHA (such as the one required for CPS projects) can also be streamlined using this data.

A lack of choice in mid haulage capacity, often leads the operators (in South Asia), to lease aircrafts with larger capacity. For new airline operations, the choice of aircrafts is not entirely based upon traffic requirements. It is also burdened by the current availability of aircraft configurations. This kind of leasing scenario is sometimes, promoted by the aircraft manufacturer, to keep the aircrafts operational (**Bjelicic, 2012**). Regulator's approval, and other commercial considerations (like the availability of aircraft fuel, at the right price), also decide, the state of airworthiness, of the current fleet of aircrafts, in regional operations.

With so many dependencies, there are less incentives for new operators, to enter air transport business. As discussed earlier in this communication, CPS activities offer a considerable advantage in operations. However, much research needs to be done, to ensure the risk-free operations. This is one of the important reasons, why, within the safety limits of operations, MEAP via CPS, needs a non-traditional approach of flight testing and approval.

In this scheme airline operators and part manufacturers can be 'ringed into' the effort as additional stake holders. Alongside the more traditional partners like academic institutes and research labs, the developmental effort can be divided into achievable limits. From a profitability perspective South Asian skies, are one of the most competitive regions of global aviation. Incentives, via tax credits, to conduct research in this non-traditional form, is one way, policy interventions can be designed. The diversion of a portion of operator's profits, into research efforts, is another form of policy intervention, which can be helpful in CPS based research on MEAP for MHA.

**Potential reuse of old aircraft and expected changes in the fuel-cell battery powered flights:**

Of the list of potential use, of old and deregistered aircrafts, research, and development- based re-use, is largely limited, to the extraction of old parts (**Mahdi Sabaghi,2016**). With proper re-engineering, a complete change in the propulsion system of an old and/or deregistered airplane, is proposed (**Schnell,2019**). This philosophy of re-use may have benefits, including that of reduction in carbon footprint.

**Issue of flight maneuverability with fuel cell-battery based power system (FCBPS).**

An in-flight aircraft, consumes fuel, at a rate, which is dependent on the flight path. Amongst all the variables, weather related disturbances in flight path, can be the hardest, to control. The flying unit needs to be maneuvered to a different zone, in case of sudden changes in weather (**Amalberti, R, 1993**). Quick maneuverability of the unit assumes importance on such occasions. This involves weight shifts, of the unit. The response time of weight shift commands, with FFPS, is much different from that of FCBPS.

To design a control mechanism FCBPS propelled flight, for a 'to-be reused' aircraft, will require a reference for the weight transfer ability. The maneuverability of such a re-engineered aircraft will be limited, in comparison to the former iteration of the same aircraft, which used fossil fuel. Gyroscopic feedback control of the weight transfer is proposed and tested for such research and development efforts (**Alex Tsai, 2009**).

Recent developments, in material and metallurgical sciences can result in a reduction in the dead-weight of the fuel cell (**Fandi Ning, 2017**) and electric-motor-generator-set (**Warwick, 2018**). The use of lightweight fuel cells is expected to improve the propulsion system, on a pre-existing, re-useable aircraft body and chassis set (**Donateo, 2017**).

With the changes in the engine and the fuel section, of the to-be re-engineered aero-plane, in place, the life of an aircraft stands extended, after the necessary approvals of flight tests. At the current state of art, the agile maneuverability of the former FFBPS enabled aircraft will be compromised in the FCBPS version. Same cannot be said about the structural fatigue. The use of electric drive reduces the vibration level of the active transporting unit. Accordingly, the rate of structural degradation, after the CPS, is expected to be lower. The two factors can be a subject matter of new research in the field of re-engineering of old aircrafts with MEAP.

The regulatory requirements of aviation sectors are strict. If these are implemented with the intention of aerosol control over regional airspaces (including South Asian skies), new test flight paths may be required. This is required for the true comparison of the data on real pollutants from air transport. This may require new testing framework over soft surfaces such as snowfields, desert beds and large water bodies using new surveying techniques.

**Conclusion**

World over, aircraft's, are assembled under licensing agreement. In South Asia, research grade micro-light & very small aircrafts have been built in private & semi-government industries. With the availability of updated data on de-registered aircrafts and associated parts, in public domain, regional aviation in South Asia has an opportunity to jumpstart the aviation manufacturing in the MHA segment. Along with support from academia and laboratories, old aircrafts, can be made available for research-based re-use. Since changes in the propulsion system, alters the flight dynamics of old aircrafts, in a major way, re-testing procedures (for airworthiness compliance) will have to evolve, in parallel with other project deadlines. All this mixed with the CPS projects for MEAP for aircrafts, is expected to revive the scenario of aircraft manufacturing related activities in South Asia.

In this paper, arguments are presented to corelate, the development of mid-haulage aviation with appropriate hybrid-energy source, via the re-engineering of flight-worthy old aircrafts. Considerations, favoring development of several mid-size airports, over a single large regional airport, in view of the development of MEAP assisted MHA are also presented. For the true development of regional aviation, the refitment exercise may turn out to be good economics. A proposal is also made for the testing of these ideas in a regional aviation sector.

This article also, proposes a modular approach, in re-assembly, where the conventional engine-based propulsion system is entirely replaced by a hybrid-electric propulsion system. Alongside the changes in the fuel systems (from fossil fuel-based liquid source to fuel cell-based battery source), feasibility of experimental research flights is also studied. It is envisioned that success in some such conversion projects, on surface (VTB$f$S) and low altitude flights (VTB$f$A), might result in validation of some of the re-engineered designs.

**(The author would like to thank the Office of Dean Research and Development (DORD) at IIT Kanpur for support).**